\documentclass[pra,twocolumn,aps,nopacs]{revtex4}

\usepackage{enumerate}
\usepackage[latin1]{inputenc}
\usepackage{amsmath}
\usepackage{amssymb}
\usepackage{mathtools}
 \usepackage{quantum}
\usepackage{color}
\usepackage[colorlinks,citecolor=blue,linkcolor=blue,linktocpage,breaklinks,hypertexnames=false,pdfpagelabels]{hyperref}
\usepackage[amsmath,thmmarks,hyperref]{ntheorem}
\usepackage[capitalise]{cleveref}

\usepackage{psfrag}
\usepackage[dvips]{graphicx}
\usepackage[all]{xy}
\newcommand{}{\`a}
\newcommand{}{\'a}
\newcommand{}{\'e}
\newcommand{}{\`e}
\newcommand{}{\'i}
\newcommand{}{\'o}
\newcommand{}{\`o}
\newcommand{}{\'u}
\newcommand{}{\"a}
\newcommand{}{\"\e}
\newcommand{}{\"\i}
\newcommand{}{\"o}
\newcommand{}{\"u}
\newcommand{Ë}{\`A}
\newcommand{}{\'E}
\newcommand{é}{\`E}
\newcommand{ê}{\'I}
\newcommand{î}{\'O}
\newcommand{ñ}{\`O}
\newcommand{ò}{\'U}
\newcommand{ì}{\"\I}
\newcommand{}{\"U}
\newcommand{}{\"A}
\newcommand{è}{\"\E}
\newcommand{ì}{\"\I}
\newcommand{}{\"O}
\newcommand{}{\"U}
\newcommand{}{\c c}
\newcommand{á}{$\cdot$}
\newcommand{}{\~n}
\newcommand{À}{?`}
\newcommand{Õ}{'}
\newcommand{á}{$\cdot$}
\newcommand{\bv}{\begin{verse}}
\newcommand{\ev}{\end{verse}}

\newcommand{\ra}{\rangle}
\newcommand{\la}{\langle}
\newcommand{\be}{\begin{eqnarray}}
\newcommand{\ee}{\end{eqnarray}}
\newcommand{\ba}{\begin{array}}
\newcommand{\ea}{\end{array}}
\newcommand{\bc}{\begin{center}}
\newcommand{\ec}{\end{center}}
\newcommand{\ben}{\begin{enumerate}}
\newcommand{\een}{\end{enumerate}}
\newcommand{\bt}{\begin{table}}
\newcommand{\et}{\end{table}}
\newcommand{\btab}{\begin{tabular}}
\newcommand{\etab}{\end{tabular}}
\newcommand{\bfi}{\begin{figure}}
\newcommand{\efi}{\end{figure}}
\newcommand{\im}{\item}
\newcommand{\bd}{\begin{description}}
\newcommand{\ed}{\end{description}}
\newcommand{\nn}{\nonumber}

\newcommand{\mc}{\mathcal}
\newcommand{\mf}{\mathbf}

\newcommand{\ts}{\textsf}
\newcommand{\trm}{\textrm}

\newcommand{\lb}{\label}

\theoremstyle{definition}

\crefname{condition}{Condition}{Conditions}
\creflabelformat{condition}{\textup{(#2#1#3)}}

\crefname{condition}{Condition}{Conditions}
\creflabelformat{condition}{\textup{(#2#1#3)}}

\theorembodyfont{\normalfont}

\theorembodyfont{\normalfont}

\theoremstyle{break}
\newtheorem{theorem-break}[theorem]{Theorem}
\newtheorem{lemma-break}[theorem]{Lemma}
\newtheorem{corollary-break}[theorem]{Corollary}
\newtheorem{definition-break}[theorem]{Definition}

\theoremstyle{nonumberplain}
\theorembodyfont{\normalfont}

\theoremsymbol{$\Box$}

\makeatletter
\def\p@subsection{}
\def\p@subsubsection{}
\makeatother

\begin{document}

\title{A Quantum Information approach to Statistical Mechanics}
\author{Gemma De las Cuevas}
\affiliation{Max Planck Institute for Quantum Optics, Hans-Kopfermann-Str.\ 1, D-85748 Garching, Germany}

\begin{abstract}
We review some connections between quantum information and statistical mechanics. 
We focus on three sets of results for classical spin models.
First, we show that the partition function of all classical spin models (including models in different dimensions, different types of many-body interactions, different symmetries, etc) can be mapped to the partition function of a single model. 
Second, we give efficient quantum algorithms to estimate the partition function of various  classical spin models, such as the Ising or the Potts model. 
The proofs of these two results 
 are based on a mapping from partition functions to quantum states and to quantum circuits, respectively.
Finally, we show how classical spin models can be used to describe certain fluctuating lattices appearing in models of discrete quantum gravity. 
\end{abstract}

\maketitle

\section{Introduction}
\lb{sec:introduction}

Quantum information and computation deals with many of the fundamental issues of quantum physics, such as non-locality or the simulatability of nature  \cite{Ni00}. 
Despite being a young research field, it has already established strong links to a number of areas, 
such as quantum optics, 
atomic and molecular physics, 
condensed matter (e.g.~in the study of strongly correlated systems), 
theoretical computer science \cite{Ar09,Bu10b}, 
or, in a smaller scale, with branches of mathematics such as operator spaces \cite{Ju10} or undecidability \cite{Va08d,Wo11c,Ei12}, 
quantum thermodynamics \cite{Re13,Br13}, 
the search of quantum effects in biology \cite{Br08b,Ca10b} 
or artificial intelligence \cite{Br12,Ma13}.
Recently, some connections between quantum information and statistical mechanics have also been established \cite{Ger08,Li97,Ge10,So07,Ve06}. 
The aim of this tutorial is to review some of these based on \cite{De11c}. This is necessarily a partial and not exhaustive review.

We will focus on classical spin models,
and prove three results from the point of view of quantum information. 
First, we will present the completeness results, 
where we show that the partition function of any classical spin model can be mapped to that of a single `complete' model. 
The proof works by first mapping partition functions to quantum states, 
and then transforming the quantum states of any model to the quantum states of the complete one.
Second, we will provide efficient quantum algorithms for estimating the partition function of various classical spin models (including the Potts and the Ising model) in a complex parameter regime. 
We also show that estimating these partition functions is \ts{BQP}-complete, i.e.~as hard as simulating arbitrary quantum computation. 
The idea of the proof is to map partition functions to quantum circuits, 
and then construct a universal gate set with the gates corresponding to a certain spin model.
Finally, we will use classical spin models to describe certain fluctuating lattices, which are of interest in certain models of discrete quantum gravity. This is a first step towards applying results in classical spin models (such as the quantum information approach presented here) to this field.

This paper is organized as follows. 
We will first introduce classical spin models (section~\ref{ssec:csm}). 
Then we will present the completeness results  (section~\ref{sec:completeness}), 
the quantum algorithms (section~\ref{sec:qalgo}), 
and the connection to discrete quantum gravity (section \ref{sec:qgravity}).
Finally we will conclude and mention some further directions in section \ref{sec:conclusions}.

\section{Classical spin models}
\lb{ssec:csm}

Most systems in nature are too complex to be described exactly:
(quantum) many-body systems, 
neurons in a brain, 
economical systems \cite{Sc03}, 
or ecosystems \cite{Me87} are just a few examples. 
Typically, variables are subject to some local optimization function (the energy,  the own economical gain, the amount of food, etc), 
and we are interested in predicting how the system behaves globally. 
The bottom-up approach of condensed matter theory consists of building a model of the system which is simple enough to handle, but rich enough to capture the relevant properties. These simplifications give rise, among others, to classical spin models.
A paradigmatic example is the Ising model \cite{Is25},  originally devised to study magnetism \cite{St71}.  
More generally, classical spin models can be used as toy models for other complex systems like the ones mentioned above. 

By a classical spin model we understand a model characterized by
(i) classical degrees of freedom, taking $q$ values, $s \in \{0,1,...,q-1\}$, and 
(ii) a cost function $H(\mathbf{s})$ depending on the configuration of the variables $\mathbf{s}$, which specifies an interaction pattern as well as the coupling strengths. 
The interaction pattern can be represented by a graph $G$  to which variables and interactions are associated in some way (as we elaborate on below), or, more generally, by a hypergraph \footnote{
A hypergraph $H$ is a pair $H=(V,E)$ where $E \ni e=\{v_{i_{1}},\ldots,v_{i_{n}}\}$, where $v_{i_{j}}\in V$. This reduces to the notion of a graph for $n=2$.
}.
Alternatively, it can be represented by a factor graph where variables and interactions are represented by different kinds of vertices \cite{Me09}.

\bfi[t]\centering
\psfrag{A}{(a)}
\psfrag{B}{(b)}
\psfrag{C}{(c)}
\includegraphics[width=1\columnwidth]{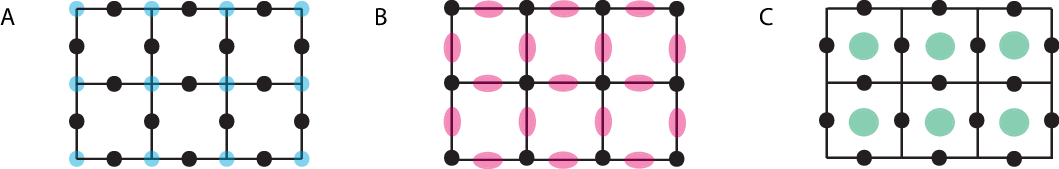}
\caption[Vertex models, edge models and lattice gauge theories]{
Different ways to associate classical spin models to graphs. 
Variables and interactions are indicated by black and colored dots, respectively.
(a) In vertex models, variables are placed in the edges, and (many-body) interactions take place in the vertices. 
(b) In edge models, variables are placed in the vertices, and (two-body) interactions take place in the edges. 
(c) In lattice gauge theories, variables sit at the edges and interactions take place on the faces. 
}
\label{fig:vertex-edge-LGT}
\efi

For interaction patterns defined on a graph $G$, various families of models can be distinguished 
depending on how particles and interactions are associated to $G$ (see Figure~\ref{fig:vertex-edge-LGT}).
In \emph{vertex models} interactions are associated to the vertices of the graph, and  variables  to the edges of the graph. Thus, interactions are (typically) many-body, and each particle can only participate in two interactions. 
For example, in a vertex model defined on a two-dimensional (2D) square lattice, each variable participates in two four-body interactions. 
Vertex models were introduced to describe ice-type models, crystals with hydrogen bonding or ferroelectrics \cite{Ba82}.

In contrast, in \emph{edge models} interactions are associated to the edges of the graph, and 
variables to the vertices.
Thus, each variable can participate in many two-body interactions. 
For example, in an edge model defined on a 2D square lattice, each variable interacts in four two-body interactions.  
Edge models were introduced to explain phase transitions in materials with elementary magnetic moments (see, e.g., \cite{Me87}). 

Finally, in \emph{lattice gauge theories} (LGTs) \cite{Ko79} variables are associated to edges and interactions to faces. 
These are usually defined on square lattices, thus interactions are always four-body.
For example, in a lattice gauge theory defined on a 2D square lattice, each variable interacts in two four-body interactions.  
LGTs were conceived, on the one hand, as gauge theories put on a lattice (the latter representing discrete spacetime \cite{Wi74}), and, on the other hand, as spin models with local symmetries, but nonetheless with a non-trivial phase diagram \cite{We71} (despite having no spontaneous symmetry breaking \cite{El75}). 
In this review we shall deal with the simplest instance of these models, namely  lattice gauge theories with gauge group $\mathbb{Z}_{2}$, called `Ising lattice gauge theories'.

In summary, the graph $G$ can be used at the convenience of the interaction pattern. 
Naturally, the models can be defined on a lattice of any dimensionality, or more generally, on an irregular graph, or a hypergraph. 
In addition, different types of interaction  (with various strengths) can take place, and models with variables with higher internal dimensions should also considered. 
This gives an idea of the wide variety of models embraced by the notion of classical spin model. 

\section{Completeness}
\lb{sec:completeness}

We have seen that classical spin models are very versatile, as they can describe systems with  different dimensionality, interaction pattern, many-body interactions, global vs.~local symmetries, etc.
Here we will show that all these models can be unified into a single  model called the `complete' model. 

To this end, we will first explain how partition functions can be related to quantum states (section~\ref{ssec:qmap}). Then we will present the main idea of the completeness proofs (section~\ref{ssec:comp}), and finally review the various completeness results (section~\ref{ssec:rev}), and discuss them (section~\ref{ssec:compconcl}).

\subsection{Relating partition functions to quantum states}
\lb{ssec:qmap}

We will first present the mapping for the Ising model, 
and finally generalize it to any model.
Consider the Ising model on a graph $G$ with vertex set $V$ and edge set $E$. 
A two-level classical spin $s_a\in\{0,1\}$ is associated to each vertex $a$, and 
this interacts with a local magnetic field $h_a$, and with neighboring spins (with inhomogeneous couplings $\{J_{ab}\}$) according to the Hamiltonian 
\be
H_G(\mathbf{s}) = -\sum_{(a,b)\in E} J_{ab} (-1)^{s_a +s_b} -\sum_{a\in V} h_a (-1)^{s_a} \, ,
\lb{eq:Ising-Hamiltonian}
\ee
where sums are performed modulo 2 throughout the section.
The canonical partition function of this model is defined as
\be
Z_{G}(J) = \sum_{\mathbf{s}} e^{-\beta H_{G}(\mathbf{s})}\, ,
\ee
where $\beta=1/(k_{B}T)$ with $k_{B}$ Boltzmann's constant and $T$ the temperature.

Our goal is to express $Z_{G}(J)$ as an inner product between two (unnormalized) quantum states \cite{Va07,Va08} (see also \cite{Hu09,Br07}). 
We first define a state $|\varphi_G\ra$ by associating each interaction to a quantum variable;  
in particular, for the Ising model defined above, we define
\be
|\varphi_G\ra :=\sum_{\mf{s}} 
\bigotimes_{(a,b)\in E} |s_a +s_b\ra  
\bigotimes_{a\in V} |s_a \ra\, .
\lb{eq:phi}
\ee
In words, this state places a quantum particle on every edge $(a,b) \in E$, 
whose state is given by the sum of the states of the classical particles sitting at the ends of that edge, $|s_a+s_b\ra$. Since the sum is modulo 2, the qubit is in the state $|0\ra$ ($|1\ra$) if the two classical particles are in the same (different) state. 
The state also places a qubit in the vertex $a\in V$ whose quantum state is the same as the classical state of the particle sitting on that vertex, $|s_a\ra$  (see Figure \ref{fig:example-Ising}). 
Finally the state sums over all configurations of classical spins. 
One can easily see that $|\varphi_G\ra$ is a stabilizer state, which is an important class of states admitting an efficient description and with many applications
\footnote{
The generators of its stabilizer group are 
$ X^{(a)} \bigotimes_{b:(a,b)\in E} X^{(a,b)}$ for all $a \in V$ and 
$Z^{(a,b)} Z^{(a)} Z^{(b)}$ for all $(a,b) \in E$, 
where we have omitted the tensor product symbols, 
$X,Y,Z$ are the Pauli matrices, 
and the superindex indicates the qubit on which the operator is acting \cite{Va07,Hu09}.}.

On the other hand, we define the product state $|\alpha\ra$ as
\be
|\alpha\ra := \bigotimes_{(a,b)\in E} |\alpha_{ab}\ra 
\bigotimes_{a\in V} |\alpha_{a}\ra \, ,
\ee
where
\begin{eqnarray}
\lb{eq:alphaab}
|\alpha_{ab}\ra &:=& e^{\beta J_{ab}}|0\ra +e^{-\beta J_{ab}}|1\ra \, ,\\
|\alpha_{a}\ra &:=& e^{\beta h_{a}}|0\ra +e^{-\beta h_{a}}|1\ra \, .
\lb{eq:alphaa}
\end{eqnarray}
That is, $|\alpha_{ab}\ra$ is a single-qubit state defined on edge $(a,b)\in E$ and, when expressed in the computational basis $\{|0\ra,|1\ra\}$, its coefficients are given by the Boltzmann weights of the interaction on that edge. 
It is immediate to see that
\be
\la \alpha_{ab}|s_a+s_b\ra = \exp(\beta J_{ab} (-1) ^{s_a+s_b})\, . 
\ee
Similarly, $|\alpha_{a}\ra$ is a single-qubit state defined on  vertex $a$ whose coefficients are given by the Boltzmann weight of the magnetic field on that vertex. It also holds that
\be
\la \alpha_{a}|s_a\ra = \exp(\beta h_a (-1) ^{s_a})\, . 
\ee
It follows that
\be
Z_G(J)=  \la \alpha (J) |\varphi_G\ra\, .
\lb{eq:mapping}
\ee

\bfi[th]\centering
\psfrag{1}{$s_1$}
\psfrag{2}{$s_2$}
\psfrag{3}{$s_3$}
\psfrag{4}{$s_4$}
\psfrag{a}{$|s_1+s_2\ra$}
\psfrag{b}{$|s_2+s_3\ra$}
\psfrag{c}{$|s_2+s_{4}\ra$}
\psfrag{A}{$|s_1\ra$}
\psfrag{B}{$|s_2\ra$}
\includegraphics[width=.5\columnwidth]{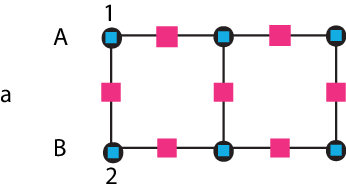}
\caption{The state $|\varphi_G\ra$ \eqref{eq:phi} (defined on a simple graph $G$) for the Ising model. It places qubits on the interactions with local magnetic fields take place (blue squares) and on the two-body interactions (pink squares).
}
\lb{fig:example-Ising}
\efi

Now we consider more general models including
 vertex models, lattice gauge theories and beyond. 
Let $I$ denote the model's interaction set, that is,  
$i\in I$ specifies the $k$ $q$-level classical spins participating in interaction $i$ (where $k$ may depend on $i$). 
Let us label their state $\mathbf{s}^{(i)}=(s_{1},\ldots, s_{k})$ (note that $\mathbf{s}^{(i)}$ is $q^{k}$-dimensional). 
Now we define $|\varphi_{G}\ra$ by associating  a quantum variable $|\mathbf{s}^{(i)}\ra$ to each interaction $i\in I$ and summing over classical variables, 
\be
|\varphi_{G}\ra = \sum_{\mathbf{s}} \bigotimes_{i\in I} |\mathbf{s}^{(i)}\ra\, .
\ee
The state $\alpha$ is defined analogously, namely
\be
&&|\alpha\ra= \bigotimes_{i\in I} |\alpha_{i}\ra\\
&&|\alpha_{i}\ra = \sum_{\mathbf{s}^{(i)}} e^{-\beta H(\mathbf{s}^{(i)})} |\mathbf{s}^{(i)}\ra\, .
\ee
It is immediate to verify that $Z_{G}(J)=\la \alpha|\varphi_{G}\ra$.

\subsection{The idea of the completeness proofs}
\lb{ssec:comp}

Now we show that the partition function of any classical spin model can be mapped to the partition function of a `complete' model. 
Consider a target model with partition function $Z_{G'}(J')$ and the another model with partition function $Z_{G}(J)$. 
To show that the latter model is complete, we need to show that, for any target, there exists a choice of couplings $J$ and a large enough $G$ such that  $Z_{G'}(J')=Z_{G}(J)$. 
To prove this, we transform the pair of states associated to $Z_{G'}(J')$  to those associated to  $Z_{G}(J)$.

More precisely,
we consider $|\varphi_{G}\ra$ defined on more variables than $|\varphi_{G'}\ra$, 
and express $|\varphi_{G'}\ra$ as the result of applying a product state $|\gamma\ra= \otimes_{i}|\gamma_{i}\ra_{i} $ to a subset of qubits of $|\varphi_{G}\ra$,
\be
|\varphi_{G'}\ra=(I\otimes \la\gamma|) |\varphi_{G}\ra\, .
\lb{eq:phiG'}
\ee
Substituting  in  $Z_{G'}(J')=\la \alpha(J')| \varphi_{G'}\ra$ we obtain
\be
Z_{G'} (J') =  (\la\alpha(J')| \otimes \la \gamma|) |\varphi_{G}\ra \, .
\ee
Since both $\alpha$ and $\gamma$ are product states, this can be rewritten as 
\be
Z_{G'} (J') =  \la\alpha(J',J'')|\varphi_{G}\ra  = Z_{G}(J',J'')\, .
\ee

In words, the partition function of the target model is written as the partition function of the complete model with couplings $J=(J',J'')$.
These are determined by the couplings of the target ($J'$), and those of the state $\gamma$ ($J''$), which specifies the interaction pattern of the target.
This shows how a target model is mapped to a certain parameter regime of the complete model.
Note that this parameter regime is provided by the construction. 
Note also that the enlargement of the complete model is precisely the size of $|\gamma\ra$, which is polynomial in all cases 
\footnote{While in Ref.~\cite{Va08} it was seen to be polynomial only for target models associated to stabiliser states,  later different ideas allowed to show that it is polynomial for all target models \cite{De09b,De10}.}

\subsection{Overview of completeness results}
\lb{ssec:rev}

We will now review four completeness results, where the complete model has been shown to be:
\begin{enumerate}[(a)]
\im The 2D Ising with fields (with complex couplings)
\lb{a}
\im The 3D Ising model (only for other Ising models)
\lb{b}
\im The 4D Ising lattice gauge theory  
\lb{c}
\im The 4D $U(1)$ lattice gauge theory for models with continuous variables
\lb{d}
\end{enumerate}
In each case, we will show why the pair of states associated to any target can be transformed to those of the complete model. 

(a) \emph{The 2D Ising with fields with complex couplings} \cite{Va08} (see \cite{Ka12b} for an alternative proof). 
Let $|\varphi_{2D}\ra$ denote the state corresponding to the 2D Ising model with fields (as defined in equation \eqref{eq:phi} with $G$ being the 2D square lattice).
Projecting each qubit associated to an edge to $\la 0_{Y}|:=\la 0|-i\la 1|$ renders the cluster state $|\mc{C}\ra$ \cite{Ra01},
\be
|\mc{C}\ra= (I\otimes \la 0_{Y}|^{\otimes |E|}) |\varphi_{2D}\ra\, .
\lb{eq:C}
\ee 
Now, the universality of $|\mc{C}\ra$ for measurement-based quantum computation guarantees that for any final state $|\varphi_{G'}\ra$ there exists a measurement pattern $\la\gamma|$ that prepares it \footnote{We are only considering one measurement branch of this measurement pattern \cite{Va08}.}, 
i.e. 
\be
|\varphi_{G'}\ra = (I\otimes \la \gamma| )|\mc{C}\ra\, .
\ee
This completes the proof. Note that these two equations imply that the couplings $J_{ij}$ are  imaginary, and that the target is encoded in the local magnetic fields $h_{i}$, which are generally complex.
Next we present completeness results with real couplings.

(b) \emph{The 3D Ising model for other Ising models} \cite{De09a}. 
Equation \eqref{eq:alphaab}  implies that the only single-qubit eigenstates of Pauli matrices that correspond to a real coupling $J'_{ab}$ are $\la 0|$ and $\la +|:=\la 0|+\la 1|$, corresponding to $J'_{ab}\to \infty$ (with proper normalization) and $J'_{ab}=0$, respectively. 
In terms of the spins, the former enforces $s_{a}$ and $s_{b}$ to be in the same state to have finite energy, and the latter decouples their interaction. 
In graph terms, the former amounts to contracting edge $(a,b)$ (called `the merge rule') and the latter to deleting it (called `the deletion rule'). 
The same conclusions apply to equation \eqref{eq:alphaa},
where $h'_{a}\to \infty$ corresponds to fixing $s_{a}=0$,
and $h'_{a}=0$ corresponds to deleting the interaction of $s_{a}$ with the magnetic field.  

Planar graphs are transformed to planar graphs under the merge and deletion rules  \cite{Di01b}. 
For this reason, we consider $|\varphi\ra$ on a non-planar graph, such as a 3D square lattice $|\varphi_{3D}\ra$ or a 2D lattice with crossings.
Then we use the merge and deletion rules to transform $|\varphi_{3D}\ra$ to $|\varphi_{\trm{clique}}\ra$ (i.e.~$|\varphi\ra$ defined on a clique or fully connected graph).
Now, since $|\varphi_{\trm{clique}}\ra$ contains all possible edges,  the state  $|\varphi_G\ra$  on any graph $G$ with  $n'\leq n$ vertices can be obtained by deleting some edges of $|\varphi_{\trm{clique}}\ra$ (defined on $n$ vertices).
This shows how to transform $|\varphi_{3D}\ra$ to  $|\varphi_G\ra$ on any $G$ by merging and deleting alone. 
However, this does not show how to transform the kind of interactions; 
in particular, if  $|\varphi_{3D}\ra$ corresponds to the 3D Ising model, $|\varphi_G\ra$ must correspond to the Ising model on a graph $G$. 
It follows that the 3D Ising model is complete with real couplings for Ising models defined on arbitrary graphs \cite{De09a}. 
Note that this is less general than \eqref{a} and, as we will see, than \eqref{c} (and \eqref{d} for continuous variables).

The merge and deletion rule have also been defined for models with $q\geq 2$-level particles and $k\geq 2$-body interactions \cite{De09a}. 
Using a similar construction as above, one concludes that a 3D $q$-level model with $k$-body interactions is complete  (with real couplings) for any other $q'\leq q$-level model with (the same kind of) $k$-body interactions.
In summary, in these results one maps models on arbitrary graphs (in particular, on arbitrary dimensions) to models with the same kind of interactions in three dimensions.
While this suggests a tradeoff between `completeness power' and real parameters, 
we show next that both features can be obtained by considering lattice gauge theories. 

(c) \emph{The 4D Ising lattice gauge theory} \cite{De09b,De10}. 
The partition function of any classical spin model (including models with any interaction pattern, kind of interactions, number of internal levels, etc.) can be recast as the partition function of an Ising lattice gauge theory on a four dimensional square lattice.
In particular, this result maps models with global symmetries (such as the Ising or Potts model), and lattice gauge theories with an Abelian and discrete gauge group, to the 4D Ising lattice gauge theory.
To prove the result, 
we define a merge and deletion rule for the 4D Ising LGT, and 
employ them to construct a superclique of Ising-type interactions. 
A superclique is an interaction pattern containing all $k$-body interactions, with  $k=0,1,\ldots, n$ (thus $2^{n}$ interactions), and Ising-type interactions are of the form $J_{i_1,i_2,\ldots, i_k} (-1)^{s_{i_1}+\ldots+s_{i_k}}$.
The main difficulty of the construction is to avoid loops of gauge-fixed variables, which is the reason why the fourth dimension is needed.
Finally we show that there exists a choice of the couplings strengths of the superclique such that this Hamiltonian equals a completely general Hamiltonian, namely one which assigns a different energy to each spin configuration.
See figure~\ref{fig:spaceofalltheories} for a representation of results
\eqref{a}, \eqref{b} and \eqref{c}. 

\begin{figure}[t]\centering
\psfrag{2}{\small{2}}
\psfrag{3}{\small{3}}
\psfrag{4}{\small{4}}
\psfrag{D}{\small{$d$}}
\psfrag{q}{\small{$q$}}
\psfrag{k}{\small{$k$}}
\psfrag{A}{\small{(a)}}
\psfrag{B}{\small{(b)}}
\psfrag{C}{\small{(c)}}
\includegraphics[width=1.1\columnwidth]{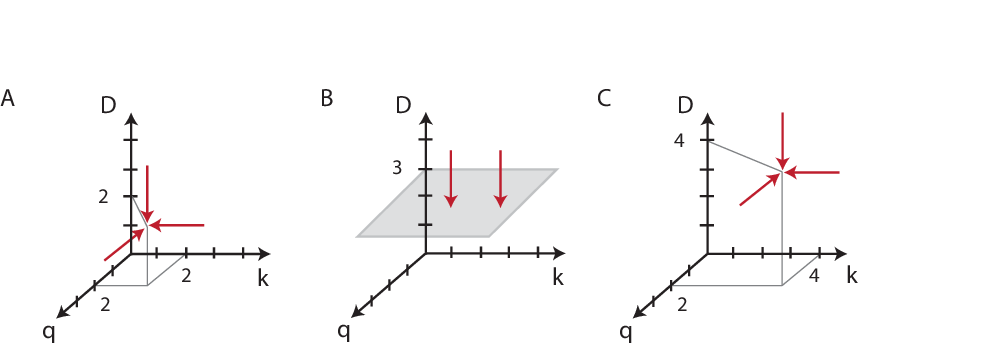}
\caption{A space representing classical spin models defined on a $d$-dimensional lattice, with $q$-level particles and with $k$-body interactions. 
(a) The completeness results show that all of these models can be mapped to the 2D Ising model with fields with complex parameters. 
(b) Also, a model on $d$ dimensions can be mapped to a 3D model with the same $q$ and $k$  and the same kind of interactions. 
(c) All classical spin models can be mapped to the 4D Ising lattice gauge theory with real couplings. 
Note that mappings to models with larger $d,q$ or $k$ are generally trivial.
}
\lb{fig:spaceofalltheories}
\end{figure}

(d) \emph{The 4D $U(1)$ lattice gauge theory} \cite{Xu11}. 
Finally, we consider classical spin models with continuous variables (i.e.~variables $x_i\in [a_i,b_i]$), and show that their partition function can be expressed approximately (but to arbitrary precision) as the partition function of the  4D  lattice gauge theory with gauge group $U(1)$. The result holds exactly for models that satisfy that 
(i) the Hamiltonian has a finite Fourier series, and (ii) that there are no constraints on the variables, and it holds approximately for models not satisfying (i) or (ii). 
Note that the result embraces models with different interaction patterns and types of many-body interactions.

To prove the result, we first generalize the quantum formulation of the partition function to continuous variable models, and define a merge and deletion rule for these models \cite{Xu11}.
Then we expand the target Hamiltonian in terms of a (truncated) Fourier series, which have the form of $\sin$ or $\cos$ of a sum of angles.
Finally, we show how to transform the 4D $U(1)$ LGT to any Fourier basis functions by means of the merge and deletion rule,
 and argue that the coefficient of each basis function can be set at will. 
Note that, as before, the couplings of the additional variables of the complete model are zero or infinite, and the other ones are real. 

Finally, we remark that completeness of $\phi^{4}$ theory has been shown in \cite{Ka12} with similar techniques. 

\subsection{Discussion}
\lb{ssec:compconcl}

The completeness results show how any classical spin model can be mapped to a specific model which is larger in size, and whose parameters specify the target model.
This means that the complete model contains all other models in its parameter regime. 
Note that generally the complete model is simpler than the target, 
thus the additional complexity is `embedded' in the inhomogeneous couplings of the additional variables of the complete model.
The enlargement of complete model is polynomial with respect to the number of parameters of the target.

\begin{figure}[t]
\centering	
\psfrag{1}{model 1}
\psfrag{2}{model 2}
\psfrag{3}{model 3}
\psfrag{4}{model 4}
\psfrag{H}{Complete model}
\psfrag{f}{fine-}
\psfrag{c}{coarse-}
\psfrag{g}{graining}
\psfrag{u}{universality}
\psfrag{6}{class 1}
\psfrag{7}{class 2}
\psfrag{N}{NP-hard}
\includegraphics[width=0.49\textwidth]{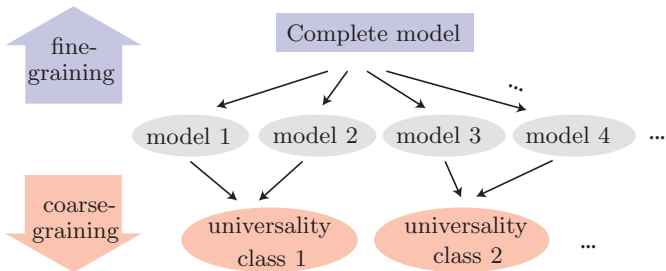}
\caption{
The completeness results provide a fine-graining procedure 
that transforms any classical spin model to (a certain parameter regime of) the complete model.
This includes models in different universality classes.
}  
\label{fig:idea-RG-tutorial} 
\end{figure}

The completeness results thus provide a fine-graining procedure that transforms any model to the complete one. 
This is the opposite of the coarse-graining procedure that Wilson proposed, which led to the classification of models into different universality classes (see figure~\ref{fig:idea-RG-tutorial}).
Note, in particular, that models of different universality classes are mapped to the complete model. 
These results provide insight into the simulation capabilities of certain classical Hamiltonians, and they have also been  used to study the simulation capabilities of quantum Hamiltonians \cite{Du11}.

Finally, we remark that the completeness results are just one application of the mapping between partition functions and quantum states presented in section \ref{ssec:qmap}. 
For example, the stabilizer state structure of $|\varphi_{G}\ra$ has been exploited to recover symmetries and high-low temperature dualities of the spin models  \cite{Va07},
 to provide efficient classical evaluations of partition functions on certain graphs \cite{Va07}, 
 and to investigate the computational capabilities of the toric code \cite{Br07} and of color code states \cite{Bo08}. 
The mapping has also been used to devise algorithms to simulate the  classical three-body Ising model \cite{Va13}. 

We also remark that this mapping is non-unique. 
Similar relations between classical systems and quantum states have allowed to relate classical and quantum phase transitions \cite{Ve06}, and  simulated annealing and quantum annealing \cite{So07}. 
In the next section we will present another mapping that will allow us to establish quantum algorithms to estimate partition functions of classical spin models.

\section{Quantum algorithms}
\lb{sec:qalgo}

Determining if the ground state of a spin model is above or below a certain energy, 
or computing its partition function are problems 
which have been traditionally studied by relating them 
to constraint satisfaction problems \cite{Mo11}. 
The \emph{quantum} computational complexity of performing these tasks has been more recently addressed \cite{Li97,Ah07}.
In this case, the relevant complexity class is \ts{BQP}, which stands for `bounded-error quantum polynomial time', and colloquially is the class of problems that can be efficiently solved by a quantum computer with a bounded probability of error \cite{Ar09}. 

Here we use a mapping between partition functions and quantum circuits established in \cite{Va09} to determine the quantum computational complexity of estimating the partition function of various classical spin models. 
We will first present this mapping (section~\ref{ssec:qcirc}), and then sketch the proof (section~\ref{ssec:bqp}).

\subsection{Relating partition functions to quantum circuits}
\lb{ssec:qcirc}

Consider the partition function $Z$ of a model defined on a lattice, so that one direction can be associated to time. 
We assume the model to have two-level variables for simplicity.
We will relate $Z$ to a quantum circuit $\mc{C}$ 
by mapping variables in the classical model to (the time evolution of) qubits in the circuit.
Additionally, each interaction will map to a quantum gate; 
more precisely, the entries of the quantum gate will be given by the Boltzmann weights of the corresponding interaction. 
Then, the product of interactions in $Z$ will map to the contraction of gates in $\mc{C}$ \cite{Va09}. 
The mapping differs slightly for vertex models, edge models and lattice gauge theories, which we present separately in the following \cite{De11}.

\begin{figure}[th]\centering
\psfrag{a}{\small{$s_1^L$}}
\psfrag{b}{\small{$s_2^L$}}
\psfrag{c}{\small{$s_3^L$}}
\psfrag{d}{\small{$s_4^L$}}
\psfrag{A}{\small{$\la s_1^L|$}}
\psfrag{B}{\small{$\la s_2^L|$}}
\psfrag{C}{\small{$\la s_3^L|$}}
\psfrag{D}{\small{$\la s_4^L|$}}
\psfrag{e}{\small{$s_2^R$}}
\psfrag{f}{\small{$s_n$}}
\psfrag{i}{\small{$i$}}
\psfrag{j}{\small{$j$}}
\psfrag{k}{\small{$k$}}
\psfrag{l}{\small{$l$}}
\psfrag{w}{\small{$w^a$}}
\psfrag{g}{\small{$\la s_1^L|$}}
\psfrag{h}{\small{$\la s_2^L|$}}
\psfrag{o}{\small{$\la s_n^L|$}}
\psfrag{u}{\small{$W^a$}}
\psfrag{1}{\small{$i$}}
\psfrag{2}{\small{$j$}}
\psfrag{3}{\small{$k$}}
\psfrag{4}{\small{$l$}}
\psfrag{p}{\small{$| s_1^R\ra$}}
\psfrag{q}{\small{$| s_2^R\ra$}}
\psfrag{r}{\small{$| s_3^R\ra$}}
\psfrag{s}{\small{$| s_4^R\ra$}}
\psfrag{P}{\small{$s_1^R$}}
\psfrag{Q}{\small{$s_2^R$}}
\psfrag{R}{\small{$s_3^R$}}
\psfrag{S}{\small{$s_4^R$}}
\psfrag{t}{\small{time}}
\hspace{-4mm}\includegraphics[width=.8\columnwidth]{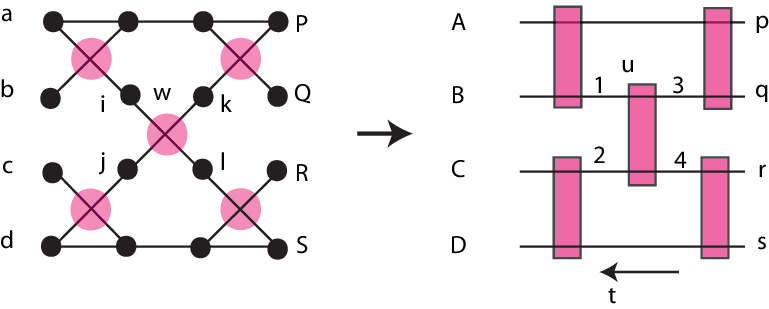}
\caption{
Mapping a vertex model (with fixed boundary conditions $(s_1^L,\ldots, s_4^L)$ on the left and  $(s_1^R,\ldots, s_4^R)$ on the right) to a quantum circuit. 
Each variable (black dots) is mapped to the time evolution of a qubit, and each interaction (pink dots) to a two--qubit gate. 
}
\label{fig:VM-tutorial}
\end{figure}

For vertex models, we consider them defined on a tilted 2D lattice  (see figure~\ref{fig:VM-tutorial}).
Vertex $a$ is associated to the four-body interaction  with Boltzmann weight $w^{a}(s_i,s_j,s_k,s_l) = \exp[-\beta h^{a}(s_i,s_j,s_k,s_l)]$. 
This interaction is mapped to a two-qubit gate $W^{a}$ as follows:
\be
W^a:=\sum_{s_i,s_j,s_k,s_l} w^a(s_i,s_j,s_k,s_l) |s_i,s_j\ra \la s_k,s_l|\, .
\lb{eq:Wa}
\ee
In words, the right (left) indices are mapped to the input (output) of the gate, and the entries of the gate are given by the Boltzmann weight of the interaction at $a$.
The corresponding quantum circuit ${\cal C}$ concatenates various layers of nearest--neighbor two--qubit gates $\mc{C} = \prod_a W^a$.

\begin{figure}[th]\centering
\psfrag{G}{\small{$\la s_1^L|$}}
\psfrag{t}{\small{time}}
\psfrag{f}{\small{$s_n$}}
\psfrag{i}{\small{$i$}}
\psfrag{j}{\small{$j$}}
\psfrag{k}{\small{$k$}}
\psfrag{l}{\small{$l$}}
\psfrag{1}{\small{$i$}}
\psfrag{2}{\small{$j$}}
\psfrag{3}{\small{$k$}}
\psfrag{4}{\small{$l$}}
\psfrag{I}{\small{$i$}}
\psfrag{J}{\small{$j$}}
\psfrag{K}{\small{$k$}}
\psfrag{m}{\small{$w^h$}}
\psfrag{n}{\small{$w^v$}}
\psfrag{5}{\small{$i$}}
\psfrag{6}{\small{$j$}}
\psfrag{7}{\small{$k$}}
\psfrag{M}{\small{$W^h$}}
\psfrag{N}{\small{$W^v$}}
\hspace{-4mm}\includegraphics[width=.6\columnwidth]{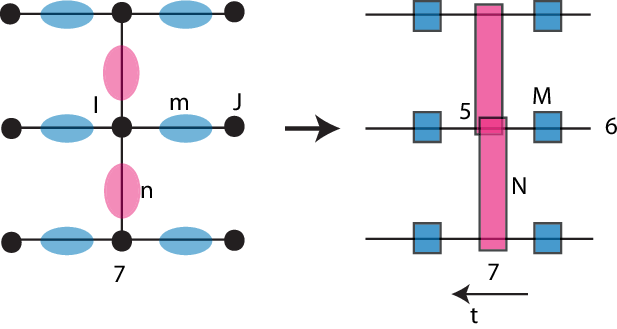}
\caption{
Mapping an edge model to a quantum circuit. 
Each variable (black dots) is mapped to (the time evolution of) a qubit. 
Each interaction along the time direction (blue ovals) is mapped to a single-qubit gate, 
and those perpendicular to time (pink ovals) are mapped to a diagonal two-qubit gate.
}
\label{fig:EM-tutorial}
\end{figure}

For edge models, we consider them defined on a 2D square lattice  (see figure~\ref{fig:EM-tutorial}). 
Edge $(a,b)$ is associated to the two-body interaction with energy  $h(s_{a},s_{b})$. 
If $(a,b)$ is along the time direction, we denote its Boltzmann weight by $w^{h}(s_{a},s_{b})=\exp[-\beta h(s_{a},s_{b})]$, and map it to a single-qubit gate, 
\be
\label{eq:W_e_horizontal} 
W^h:= \sum_{s_i, s_j} w^h(s_i, s_j) |s_i\rangle\langle s_j|\, .
\ee 
If $(a,b)$ is perpendicular to time, we denote its Boltzmann weight by $w^{v}$ and map it to a diagonal two-qubit gate, 
\be
\label{eq:W_e_vertical} 
W^v:= \sum_{s_i, s_k} w^v(s_i, s_k) |s_i, s_k\rangle\langle s_i, s_k|\, .
\ee 
The circuit ${\cal C}$ consists of alternating layers of operations associated with the horizontal and vertical edges of the 2D lattice ${\cal C}=\prod_{h,v}W^hW^v$. 

\begin{figure}[th]\centering
\psfrag{w}{\small{$w^t$}}
\psfrag{W}{\small{$W^t$}}
\psfrag{S}{\small{$W^s$}}
\psfrag{s}{\small{$w^s$}}
\psfrag{t}{\small{time}}
\psfrag{1}{\small{$i$}}
\psfrag{2}{\small{$j$}}
\psfrag{3}{\small{$k$}}
\psfrag{4}{\small{$l$}}
\psfrag{5}{\small{$m$}}
\hspace{-4mm}\includegraphics[width=.95\columnwidth]{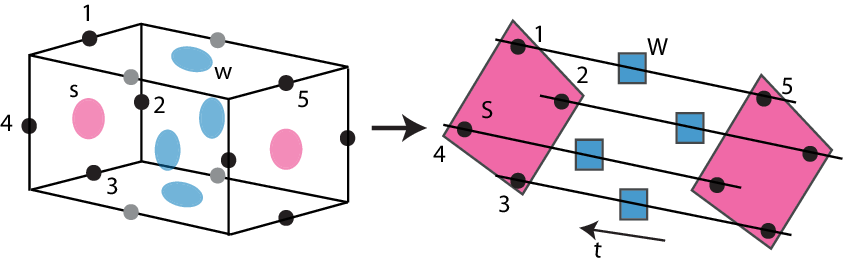}
\caption{
Mapping a 3D LGT in the temporal gauge to a quantum circuit. 
Each variable (black dots) is mapped to the time evolution of a qubit,
interactions along the time direction (blue ovals) are mapped to single-qubit gates, 
and those perpendicular to to time (pink ovals) to four-qubit diagonal gates.
}
\label{fig:LGT-tutorial}
\end{figure}

Finally, we consider a 3D Ising lattice gauge theory in the temporal gauge \footnote{The temporal gauge is obtained by fixing all variables (that sit on the edges) in the time direction.} (see figure~\ref{fig:LGT-tutorial}).
Face $f$ along the time direction is associated to the two-body interaction with Boltzmann weight 
$w^t (s_i,s_m) := \exp[-\beta h_f(s_i,s_m)]$. This is mapped to a single-qubit gate $W^t$, 
\be
W^t : =\sum_{s_i,s_m} w^t (s_i,s_m) |s_i\ra \la s_m|\, .
\lb{eq:singlequbitgate}
\ee
In addition, face $f$ perpendicular to time is associated to a four-body interaction with  Boltzmann weight 
$w^{p}( s_i,s_j,s_k,s_l)=\exp[-\beta h(s_i,s_j,s_{k},s_{l})]$.
This is mapped to a four-qubit diagonal gate $W^p$, 
\be
W^p &:=& \sum_{s_i,s_j,s_k,s_l} 
w^p (s_i,s_j,s_k,s_l)\times \nn\\
&&|s_i, s_j, s_k, s_l\ra \la s_i, s_j, s_k, s_l |\, .
\label{eq:F}
\ee
The quantum circuit takes the form $\mc{C} = \prod_{p,t} W^p W^t$.

Finally, if the spin model has fixed boundary conditions on the right $R=(s_1^R,\ldots,s_n^R)$ and on the left $L=(s_1^L,\ldots,s_n^L)$,
 these are mapped to the input $|R\ra=|s_1^R\ra\otimes \ldots \otimes |s_n^R\ra$ and output $|L\ra=|s_1^L\ra\otimes \ldots \otimes |s_n^L\ra$ of the circuit, resulting in
\be
Z= \la L|\mc{C}|R\ra\, ,
\lb{eq:mapping-qcircuit}
\ee
whereas for open boundary conditions it holds that 
$Z = \la +|^{\otimes n} \mc{C}|+\ra^{n}$, 
 and for periodic boundary conditions, $Z = \tr(\mc{C})$ \cite{Va09,De11}. 

\subsection{\ts{BQP}-completeness of classical spin models}
\lb{ssec:bqp}

Now we use the mapping \eqref{eq:mapping-qcircuit} to determine the quantum computational complexity of estimating $Z$. 
The idea is to relate the estimation (with an additive approximation) of $\la L |\mc{C}|R \ra$ to the estimation of the matrix element $\la 0 |^{\otimes n} U |0\ra^{\otimes n}$, where $U$ is a unitary consisting of a polynomial number of two-qubit gates. 
Since $U$ can be expressed (up to polynomial accuracy) using polynomially many gates from a universal gate set  \cite{Ni00}, 
the goal is to construct a universal gate set with gates corresponding to a certain classical model.
If this is possible, then estimating $\la L |\mc{C}|R \ra$ 
is as hard as estimating $\la 0 |^{\otimes n} U |0\ra^{\otimes n}$, which is easily seen to be a \ts{BQP}-complete problem using the Hadamard test \cite{De11}.
It follows that estimating $Z$ (in the parameter regime leading to the universal gate set)
 is \ts{BQP}-complete. 
Moreover, the proof is constructive, 
as it provides the pattern of couplings of the spin model 
that encodes the quantum algorithm to estimate $Z$ (in that parameter regime). 

In summary, 
the goal is to show that for a given classical model 
there exists a choice of coupling strengths such that the corresponding gates are 
(i) unitary and 
(ii) form a universal gate set. 
For example, for the 2D Ising model we need to see if there is a coupling strength $J_{ij}$ at edge $(i,j)$ such that the corresponding gate $W$ is unitary and realizes a particular gate of a universal gate set. 
Note that in general (i)  implies that the coupling strengths are imaginary, 
since only gates with positive entries correspond to real Boltzmann weights, and thus real coupling strengths
\footnote{While real universal gate sets have been identified \cite{Sh03}, to the best of our knowledge 
no real and positive universal gate set has been found.} 
(see \cite{Ib12} for similar results).
Moreover, in most cases several gates (each corresponding to a certain coupling strength) must be considered together so that their overall action is like that of a gate from the universal gate set. 
This implies that this arrangement of coupling strengths must appear together for the result to hold.

This way, we show that estimating the partition function of 
(1) the six vertex model on a 2D square lattice,
(2) the Ising model with magnetic fields on a planar graph, 
(3) the Potts model on a quasi 2D square lattice
\footnote{That is, a 2D square lattice surrounded by qubits in a fixed state, see figure 8 in \cite{De11}.}, and 
(4) the 3D Ising lattice gauge theory, 
all in a certain complex parameter regime, is \ts{BQP}-complete, and we provide the efficient quantum algorithms to estimate them \cite{De11}.

\section{Discrete quantum gravity}
\lb{sec:qgravity}

So far, our results have applied to classical spin models with a fixed coupling and interaction pattern.
For example, a model with a fixed, but arbitrary, coupling and interaction pattern can be mapped to the complete model. 
Yet, classical spin models with random coupling strengths (also called with `bond disorder') are used in spin glasses \cite{Me87} and in fault-tolerant quantum computation \cite{De02}, 
while models on random graphs  (also called with `connectivity disorder')
are used as toy models of matter in some discrete quantum gravity models, 
such as causal dynamical triangulation (CDT)  \cite{Am00,Am09a}. 

Here we focus on the latter class of models, and in particular on the case without matter.
This theory follows Feynman's approach for quantization, 
where the quantity of interest  is the transition amplitude between an initial and a final state. 
After discretizing spacetime, and performing a rotation to imaginary time, this quantity takes the form of a partition function, namely $\sum_T e^{-S_T}$ where $S_{T}$ is the action of the gravitational field \footnote{In CDT, Regge's discretization of the Einstein--Hilbert action is used.} and $T$ is a discretized space-time, which can be treated as a (weighted) graph. The latter is also called a fluctuating lattice.

We attempt to use classical spin models to define the fluctuating lattice itself. 
This may allow one to export results of spin models (including the quantum information approach presented here) to the field of discrete quantum gravity. 
We have accomplished a small step in this direction, namely, 
we have described discretized space-time in 1+1 dimensions without matter (as conceived by CDT) 
 in terms of independent classical degrees of freedom, which can be seen as spins  \cite{We12}.

More precisely, we consider two-dimensional triangulation with a global proper time, also called `foliated triangulations'. 
The central observation is that, while the building block of a triangulation is the triangle, 
the building block of a foliated triangulation is a pair of triangles that share a space-like edge. 
We call this basic unit a `fork', as it consists of a vertex with three edges (the middle edge being at constant time), together with the two faces lying between these edges (see figure~\ref{fig:forks}). 
This implies that any  foliated triangulation  $T$ can be assembled entirely out of forks.

\bfi[t]\centering
\psfrag{A}{(a)}
\psfrag{B}{(b)}
\psfrag{C}{(c)}
\psfrag{t}{\small{$t$}}
\psfrag{t+1}{\small{$t+1$}}
\psfrag{t+2}{\small{$t+2$}}
\psfrag{t+3}{\small{$t+3$}}
\includegraphics[width=1\columnwidth]{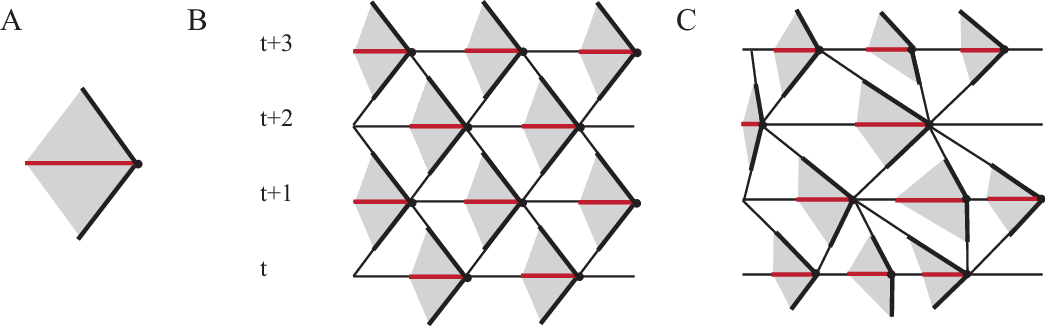}
\caption{(a) The basic building block of a foliated triangulation, the `fork', consists of a vertex, 3 edges (the red one at constant time), and 2 faces (marked in gray).
(b) The foliated triangulation with all forks present corresponds to the triangular lattice. 
(c) An arbitrary, foliated triangulation $T$ assembled out of forks.}
\lb{fig:forks}
\efi

Thus, to describe $T$ we only need to specify the order in which the forks are assembled.
Equivalently, we fix an order to assemble the forks (bottom to top, and left to right), and specify at each step (labeled by the vertical coordinate $n$ and the horizontal one $m$) whether the fork is assembled or not. 
More precisely, we associate a binary variable $\lambda_{nm}$ describing the presence ($\lambda_{nm}=1$) or absence (0) of the fork at position $n,m$. 
For example, the foliated triangulation with all forks present  ($\lambda_{nm}=1$ for all $n,m$) corresponds to the triangular lattice  (see figure~\ref{fig:forks}). 
This shows how to associate an arbitrary bit array $\{\lambda_{nm}\}_{n,m}$ to a foliated triangulation  (see Figure~\ref{fig:forks-2}). For the converse, see \cite{We12}. 

This binary description allows us to rephrase various quantities of CDT (such as the Ricci scalar, the action or the volume) in binary language \cite{We12}.
Note that this description is in spirit similar to that of a lattice gas model, where a fluid (with molecules absent or present) is described in terms of a magnet with two-level spins on a fixed lattice \cite{St71}.

\bfi[t]\centering
\psfrag{0}{0}
\psfrag{1}{1}
\includegraphics[width=1\columnwidth]{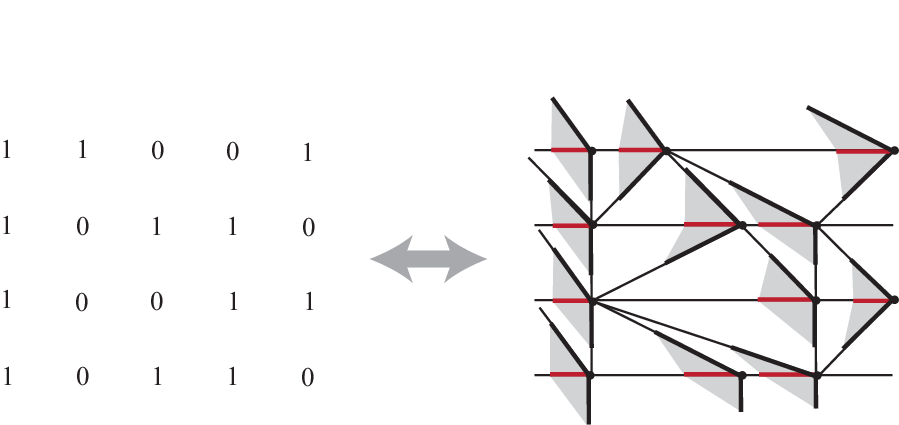}
\caption{
A two-dimensional bit array specifies a foliated triangulation
by interpreting each 0 (1) as the absence (presence) of a fork, and fixing an order to assemble the forks. 
}
\lb{fig:forks-2}
\efi

\section{Conclusions and outlook}
\lb{sec:conclusions}

We have reviewed an approach to statistical mechanical problems from the point of view of quantum information based on  \cite{De11c}.
We have presented three results. 
First, we have shown the completeness results, where the partition function of all classical spin models is  mapped to the partition function of a single model. 
The complete models have been shown to be the 2D Ising model with fields with complex couplings, 
the 3D Ising lattice gauge theory for other Ising models,
the 4D Ising lattice gauge theory, and the 4D $U(1)$ lattice gauge theory for models with continuous variables.
The idea of the proof is to map partition functions to quantum states, and to transform  quantum states of a model to those of another.

Second, we have shown that estimating the partition function (with an additive approximation) of various classical spin models (in a complex parameter regime) is \ts{BQP}-complete, i.e.~it is as hard as simulating arbitrary quantum computation. 
These models include the 2D six vertex model, the Ising model with magnetic fields on a planar graph, the  Potts model on a quasi 2D lattice, and the 3D Ising lattice gauge theory.
The idea of the proof is to map partition functions to quantum circuits, and construct a universal gate set with the gates corresponding to a certain spin model.

Finally, we have described 2D foliated triangulations in terms of a classical spin model. These lattices are seen as a discretization of two-dimensional space-time without matter in a theory of discrete quantum gravity called causal dynamical triangulation. 
This may allow to apply spin model results to this field. 

Concerning the completeness results, we have very recently  defined a notion of completeness for Hamiltonians (which implies completeness for partition functions), and have provided sufficient and necessary conditions for a model to be Hamiltonian complete \cite{De13}. This allows us to prove that the 2D Ising model with fields is complete with real couplings. 
It would be very exciting to extend these results to quantum Hamiltonians.

Concerning the complexity results, it would be very interesting to prove them in a real parameter regime (which is the one used in most applications of classical spin models), or in the parameter regime and accuracy scale for which classical results known, so that the classical and quantum computation complexity of this problem can be compared.
 
 Finally, it would be very interesting to see if our spin formulation of 2D foliated triangulations can be extended to 3 or 4 dimensions, and if our quantum approach can lead to new insights. 

\emph{Acknowledgements}.
I would like to thank my colleagues M. Van den Nest, W.~D\"ur, H. J. Briegel and M. A. Martin--Delgado for their help and work during all these years. 
I also thank the Alexander von Humboldt foundation for support.

\bibliographystyle{apsrev}

\bibliography{/Users/gemmadelascuevas/Dropbox/Gemma/Special-files/all-my-bibliography.bib}

\end{document}